\documentclass[11pt]{article}
\usepackage{amsmath,amsthm,amssymb}
\usepackage{graphicx}
\usepackage{psfrag}

\usepackage[letterpaper,margin=0.75in]{geometry}

\def\R{\mathbb{R}}
\def\C{\mathbb{C}}

\def\P{\mathbb{P}}

\def\P{\mathbb{P}}

\newcommand{\be}{\begin{equation}}
\newcommand{\ee}{\end{equation}}
\newcommand{\bea}{\begin{eqnarray}}
\newcommand{\eea}{\end{eqnarray}}
\newcommand{\beann}{\begin{eqnarray*}}
\newcommand{\eeann}{\end{eqnarray*}}
\newcommand{\benn}{\begin{equation*}}
\newcommand{\eenn}{\end{equation*}}

\def\ra{\rightarrow}
\def\I{\infty}
\def\I{\infty}

\begin{document}

\author{Christian Kuehn\footnotemark[1]~ and Gerd Zschaler\footnotemark[2]~ 
and Thilo Gross\footnotemark[3]}

\renewcommand{\thefootnote}{\fnsymbol{footnote}}
\footnotetext[1]{%
Institute for Analysis and Scientific Computing, 
Vienna University of Technology, 
1040 Vienna, Austria.} 
\renewcommand{\thefootnote}{\arabic{footnote}}

\renewcommand{\thefootnote}{\fnsymbol{footnote}}
\footnotetext[2]{%
TNG Technology Consulting, 
85774 Unterf{\"o}hring, Germany.} 
\renewcommand{\thefootnote}{\arabic{footnote}}

\renewcommand{\thefootnote}{\fnsymbol{footnote}}
\footnotetext[3]{%
University of Bristol, 
Merchant Venturers School of Engineering, 
BS8 1TR Bristol, UK.} 
\renewcommand{\thefootnote}{\arabic{footnote}}

\renewcommand{\thefootnote}{\fnsymbol{footnote}}
\footnotetext[4]{%
Max Planck Institute for the Physics of Complex Systems (MPI-PKS),
01187 Dresden, Germany.
(The majority of this work was carried out while all three authors worked at the MPI-PKS; hence
we would like to acknowledge it here for hospitality and financial support).} 
\renewcommand{\thefootnote}{\arabic{footnote}}

\title{Early warning signs for saddle-escape transitions\\ 
in complex networks}

\maketitle

\begin{abstract}
Many real world systems are at risk of undergoing critical transitions, leading to sudden qualitative and sometimes irreversible regime shifts. The development of early warning signals is recognized as a major challenge. Recent progress builds on a mathematical framework in which a real-world system is described by a low-dimensional equation system with a small number of key variables, where the critical transition often corresponds to a bifurcation. Here we show that in high-dimensional systems, containing many variables, we frequently encounter an additional non-bifurcative saddle-type mechanism leading to critical transitions. This generic class of transitions has been missed in the search for early-warnings up to now. In fact, the saddle-type mechanism also applies to low-dimensional systems with saddle-dynamics. Near a saddle a system moves slowly and the state may be perceived as stable over substantial time periods. We develop an early warning sign for the saddle-type transition. We illustrate our 
results in two network models and epidemiological data. This work thus establishes a connection from critical transitions to networks and an early warning sign for a new type of critical transition. In complex models and big data we anticipate that saddle-transitions will be encountered frequently in the future.  
\end{abstract}

The low-dimensional systems that are commonly investigated in the context of critical transitions 
(or tipping points)\cite{Schefferetal,Lentonetal} typically settle to stable, but not necessarily stationary,  
states\cite{KuehnCT1}. The defining features of such states is that the system returns exponentially to 
the state after sufficiently small perturbations\cite{GH}. When environmental parameters change, a critical 
transition may occur when thresholds are crossed, where the system becomes unstable to certain 
perturbations. Low dimensional systems then typically depart quickly from the original state before 
approaching another, possibly distant, state. The thresholds at which such transitions occur are called 
\emph{bifurcation points}\cite{Kuznetsov} and the corresponding transitions are \emph{local bifurcations}
\cite{AshwinWieczorekVitoloCox,KuehnCT1}. Warning signs for such bifurcation-induced transitions detect the 
loss of exponential restoring dynamics either through its impact on the statistics of noise-induced 
fluctuations\cite{Schefferetal,CarpenterBrock,BoettingerRossHastings} or by direct measurement of recovery 
rates\cite{LadeGross,Veraartetal}. Thus, bifurcation-induced critical transitions are well understood, and 
the corresponding warning signs have been analyzed mathematically\cite{KuehnCT2} and tested in 
experiments\cite{DrakeGriffen,DaiVorselenKorolevGore}.
 
For understanding the alternative mechanism of critical transitions that is the focus of the present paper, 
consider that many real-world complex systems do not reside in stable states. An intuitive example is 
provided by the outbreak of an epidemic invasion in a population occuring under stable enviromental conditions. Although 
the precise mechanism for spreading after invasion can be very complex\cite{DushoffLevin}, many examples 
show that introducing a certain new or previously extinct pathogen into an unprepared population 
can have drastic consequences. This illustrates that the original state of the system, in 
which the pathogen is absent, was unstable with respect to a specific perturbation, corresponding 
to the introduction of the pathogen. Note that even after its introduction, the pathogen may be extremely 
rare for an extended amount of time, residing in small subpopulations or animal vectors, such that 
on a macroscopic level a disease-free state is still observed for significant time. We may only see a very 
small rise in the number of infected individuals for a long time under fixed environmental conditions 
but eventually a drastic jump to an endemic regime occurs.

\begin{figure}[htb]\centering
\includegraphics[width=1\textwidth]{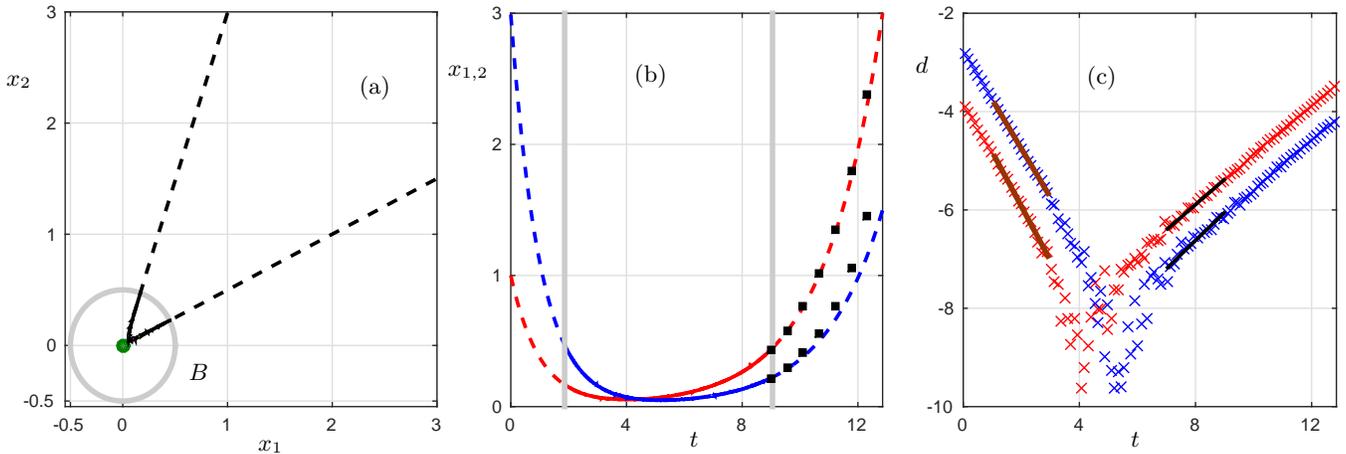}
\caption{\label{fig:fig1}Dynamics near a planar saddle with small noise. (a)
Phase space $(x_1,x_2)$ with a trajectory (black) passing near the saddle point.
The stationary state (dark green dot) and a circle (gray) of radius $r=0.5$ indicate
a neighborhood of the stationary state outside of which the trajectory is shown as
a dashed curve. (b) Time series for $x_1$ (red) and $x_2$ (blue). The gray
vertical lines indicate entry and exit to the ball
$B=\{x\in\R^2:x_1^2+x_2^2<r^2\}$. The black squares are predicted values from
the warning signals obtained inside $B$. (c) Plot of the logarithmic
distance reduction $d(T)$ as crosses; (see Supplementary Information, Section 2). The
red/blue linear interpolants yield two approximations for the stable eigenvalue
$\lambda_s\approx -1.10,-0.99$ and the black lines for the important unstable
eigenvalue $\lambda_u\approx 0.51,0.57$; the true values are
$(\lambda_s,\lambda_u)=(-1,0.5)$. The black squares in (b) can be obtained from
$x_{1,2}\sim e^{\lambda_u t}$. Note that the choice of $B$ is a choice of sliding
window length (or lead time) for prediction as in the case for bifurcation-induced 
tipping.}
\end{figure}

The example above differs fundamentally from bifurcation-induced transitions, because the qualitative change 
is not induced by a change of environmental parameters, but rather by a specific `rare' perturbation. A system 
is susceptible to such perturbation-induced transitions if it resides in a saddle point, a state that is stable with 
respect to some perturbations, but unstable with respect to others. In the following we refer to critical
transitions caused by the departure from saddle points far from bifurcations as \emph{saddle-escape transitions};
see also (Fig.~\ref{fig:fig1}) for a mathematical normal form example for passage near a saddle; we remark 
that this example is generic in the sense that
mathematical theory guarantees that other systems with nondegenerate saddles show the same
dynamics up to coordinate changes and by using a suitable notion of equivalence for the dynamics.   
       
In simple low-dimensional, modular or symmetric ({i.e.} effective few-variable) systems that are typically 
studied in the context of early-warnings signs, saddle-escape transitions may 
occur\cite{CushingDennisDesharnaisCostantino,Hastings7} but are relatively rare in practice, particularly 
in comparison to saddles in high-dimensional (many-variable) systems. To understand this difference, first, 
consider the abundance of available saddle states. Whether a steady state is an attractor, repeller, or 
saddle is determined by the eigenvalues of the systems Jacobian matrix, which provides a linearization 
of the system around the state in question\cite{GH,Kuznetsov}. A state is a repeller when all of these 
eigenvalues have positive real parts, an attractor if all eigenvalues have negative real parts, and a 
saddle when there are eigenvalues with positive real part and also eigenvalues with negative real-part. In a complex heterogeneous system the 
eigenvalues can, to first approximation, be considered as random variables\cite{May1} that have negative 
real parts with a certain probability $q$. Since the total number of eigenvalues increases with the 
number of variables $N$, the proportion of attractors decreases as $q^N$, the proportion of repellers 
decreases as $(1-q)^N$, whereas the proportion of saddles increases rapidly with increasing system size 
(see Supplementary Information, Section 1). We can therefore expect to find an abundance of saddle states 
in generic complex high-dimensional heterogeneous systems.
 
Given the existence of saddles we may also want to to ask whether systems can even 
approach a single saddle. This is possible, {e.g.}, if a trajectory exists 
that connects the saddle state to itself or several trajectories connect between different saddles. Such 
\emph{homoclinic} or \emph{heteroclinic orbits}\cite{GH,Kuznetsov} exist in low-dimensional systems 
only if certain conditions are met exactly, e.g.~parameters are tuned exactly right. However, if we 
allow parameters to change dynamically, the dimension of the system increases naturally and long-time 
homoclinic and heteroclinic dynamics can appear robustly (see Supplementary Information, 
Section \ref{sec:saddle_warning}, for an epidemic model example). A simple example is the fold-homoclinic 
(or square-wave) bursting mechanism observed in many neurons\cite{Izhikevich}, where additional 
macroscopic slow gating variables are frequently introduced to capture the complex processes in
the neuron in addition to the usual voltage dynamics. This leads to a robust repeated passage near 
a homoclinic structure. Furthermore, if there are certain symmetries in the system, then heteroclinic 
behavior may occur generically as well\cite{AshwinField,HuismanWeissing}.     
 
Saddle-escape can only be considered a critical transition if the system resides in the saddle for 
a sufficiently long time such that the saddle is perceived as the natural state. The residence time of a system 
near a saddle can be long as the system moves slowly near steady states (see Supplementary 
Information, Section 1, for a standard calculation of the residence time). When the system is subject 
to a small perturbation, its response can be decomposed into different fundamental modes. Some of these
modes decay quickly, restoring the system to the steady state. However, at least one mode exists that 
once excited initially grows exponentially, and thus leads to an escape from the saddle. Here we consider 
the state where the initial perturbations from those growing modes are very small or entirely absent, 
since otherwise the system would depart very quickly from the saddle. 

In full mathematical generality we would usually expect that a typical perturbation of the system excites all 
modes to some degree, and thus triggers departure from the saddle point. Based on this reasoning we would expect that already the first 
perturbation of a system residing in a saddle launches it into exponential departure. However, in 
real high-dimensional systems the situation is not so simple. For example, hidden symmetries or constraints may exist that 
effectively decouple certain variables from perturbation of other variables. The most important constraint is the absolute 
nature of the zero line. Consider again the example of a rare pathogen introduced into a population. Here it is immediately 
apparent that a perturbation cannot excite the exponential growth of the infected population, unless it already involves 
the introduction of at least some infected individuals. Furthermore, the perturbation direction to the saddle may initially be very weak in
comparison to the strong stable directions, which also leads to long residence times near saddle points (see
Supplementary Information, Section \ref{sec:saddle_warning}, for a simple compartmental epidemic model illustrating
this effect). Finally, in many 
models stochastic effects cannot be ignored close to the steady state (see Supplementary Information, Section 2). For 
instance it is well known that small populations that are feasible in the deterministic limit may still go extinct due 
to fluctuations in populations size, a phenomenon known as demographic extinction\cite{LiebholdBascompte}. These mechanisms, {i.e.}, 
micro-level stochastic extinction, slowness of departure and decoupling of perturbation modes, can effectively stabilize 
the saddle state for a long time until a particular perturbation, or series of 
perturbations, launches the system on an escaping trajectory. Considered together, the well known-arguments 
presented above suggest that saddle-escape may play a role as a critical transition, particularly in 
high-dimensional systems. 

Let us now ask whether warning signs for saddle transitions can exist. Because saddle-escape is triggered 
by the occurrence of a certain perturbation, which is inherently unpredictable in the context of the model, we 
cannot hope to detect the saddle escape far before the exponential departure from the saddle starts. However, 
consider that in contrast to bifurcation-induced transitions, saddle escape will generally occur at states 
that are far from bifurcations. Such states are said to be \emph{hyperbolic}, and perturbations to these 
states grow or decline exponentially depending upon the perturbation direction. The departure form the saddle is initially 
slow and dynamics around the saddle are often perceived as meta-stable. Phenomenologically, the transition 
therefore appears very similar to a bifurcation-induced critical transition, showing first a slow drift, followed 
by a sharp spike. Arguably, in many applications it should thus be possible to restore the system to the saddle 
during the initial phase where the intrinsic dynamics are still slow. Our aim is thus to detect the saddle-escape 
after the critical perturbation has occurred, but before the system has moved so far from the saddle that the dynamics 
has accelerated too much. In fact, if we know, a priori, the system is approaching a saddle, 
not just a fully stable state, then the warning sign described below may help to detect a potential 
critical transition already at the very beginning of the metastable phase; otherwise, we can apply 
it during the metastable phase.     

In our mathematical treatment we consider a scenario where the system starts from some initial point, 
then approaches the saddle and stays for a significant time in the vicinity of the saddle point 
before departing again. For detecting the onset of the departure we exploit the exponential form 
of the departure trajectory. Between two time points $t_1$ and $t_2$ the logarithmic distance along 
a trajectory close to the saddle point $x^*$ will eventually be dominated by a scaling of the form
\be
\label{eq:est_u_main}
\ln \|x(t_2)-x(t_1)\|\approx \lambda_u t_2+k_0,
\ee
where $\lambda_u>0$ is the real part of the largest eigenvalue of the Jacobian and $k_0$ is a 
constant (see Supplementary Information, Section 2), {i.e.}, the logarithmic distance
increases linearly in forward time. Although saddles generically have positive eigenvalues, 
the influence of $\lambda_u$, which can be estimated by Eq.~\ref{eq:est_u_main}, on the overall 
system dynamics is small as long as the system is approaching the saddle sufficiently close to 
its stable directions. In this case, the dynamics does not yet involve a significant component in the 
direction of unstable eigenvectors or the system would not approach the saddle at all. 

Let us illustrate this again by the epidemics example. While the pathogen is absent the variable 
that captures the density of the infected population remains fixed to zero, since there are no 
dynamics in the infected population variable we only measure a stable eigenvalue with real part 
$\lambda_s<0$ via an analogous logarithmic scaling relation as Eq.~\eqref{eq:est_u_main} (see 
Supplementary Information, Section 2), where the logarithmic distance decays linearly in forward 
time. Only after the pathogen is introduced and the infected population starts growing, we 
start to pick up the positive eigenvalue $\lambda_u$ associated to the exponential growth.   

Hence, we may use logarithmic distances between points as a measure to determine, 
which eigenvalue $\lambda$ is currently dominating as shown in (Fig.~\ref{fig:fig1} (b)--(c)).
If we find negative $\lambda$ this signals that we are in a regime dominated by a stable 
direction and approach the saddle, whereas a positive $\lambda$ signals a departure. Therefore, 
the emergence of positive $\lambda$ beyond a certain threshold can provide a warning signal 
for saddle-escape transitions. If we know, a priori, that we are approaching a saddle, not just a fully
stable state, then the warning sign already detects the saddle when the logarithmic distance
reduction decays linearly. If it could be a stable or saddle state, we can only apply the logarithmic
distance reduction warning sign during the metastable phase near the saddle.     

To illustrate the saddle-escape mechanism for critical transitions and to test the proposed 
early-warning signal we consider two recent adaptive-network models in which critical transitions 
as well as saddle-type behavior play key roles in the dynamics and an epidemiological data set. The 
first system is a model from evolutionary game theory, in which the evolution of cooperative 
behavior in a network of interacting, self-interested agents is studied\cite{ZschalerTraulsenGross}. 

The model describes a network of agents, connected by social contacts.
Each agent pursues one of two possible strategies, which we call cooperate and defect.  
The agents engage in pairwise interactions with their neighbors, which are modeled as 
a snowdrift game\cite{DoebliHauert}. In this game the highest social payoff in produced by 
mutual cooperation. However, for the individual agent, defecting yields a higher payoff
when the agent is interacting with a cooperator.     
In time both the agents' strategies and the network of interactions change as agents switch to the 
strategy that performs optimally in the population, and also rewire their connections to other agents 
following the more successful strategy (see Supplementary Information, Section 3).

Previous work\cite{ZschalerTraulsenGross} has shown that in the limit of infinite population 
size the system robustly approaches the state of full cooperation, where the probability that 
a randomly drawn agent follows the cooperative strategy is one. In large, but finite, networks a 
state of almost full cooperation is reached that is disrupted by large outbreaks of defection 
(see Supplementary Information, Section 3).

\begin{figure}[htb]\centering
\includegraphics[width=1\textwidth]{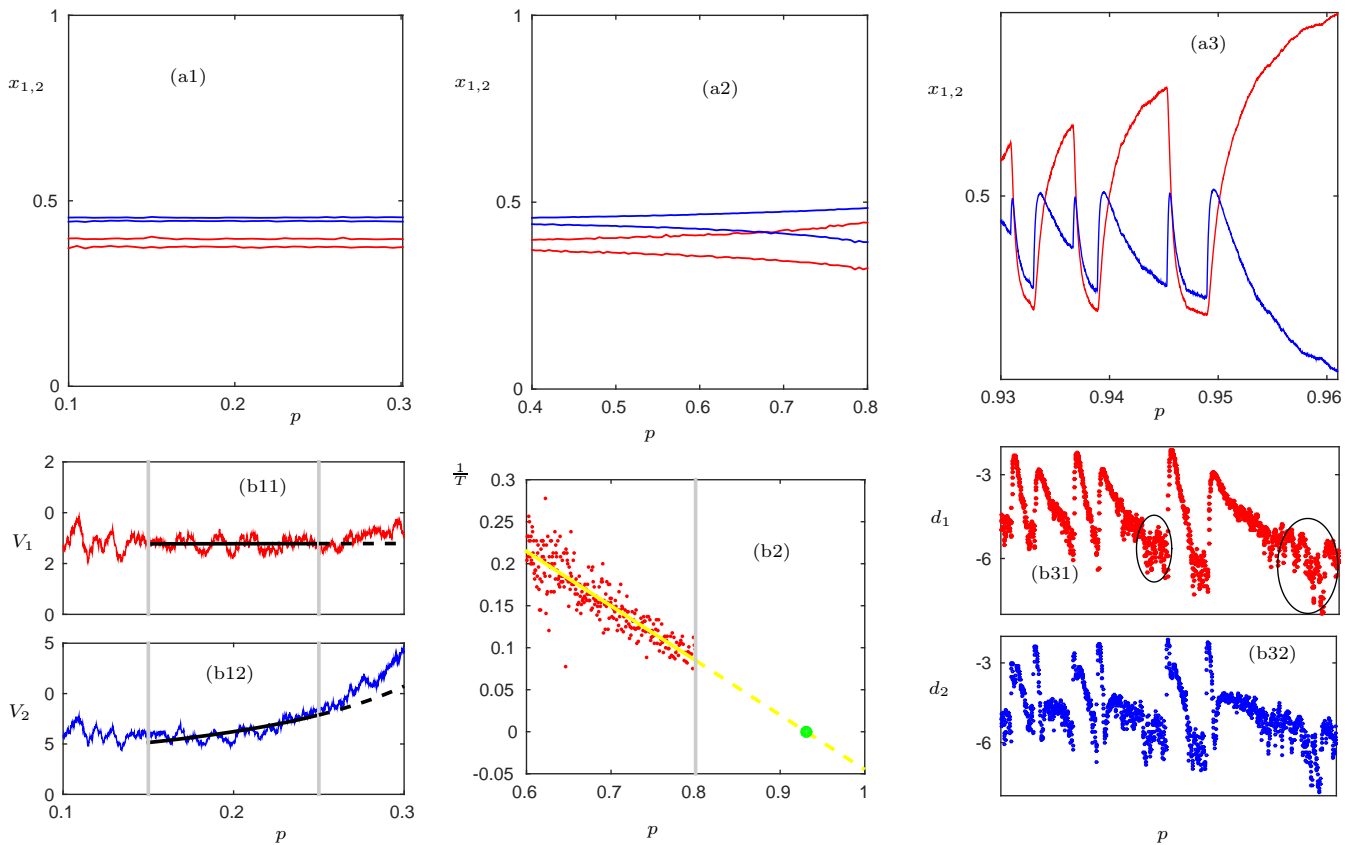}
\caption{\label{fig:fig2}Critical transitions for an evolutionary game. (a) Time
series for the density of cooperators $x_1$ (red) and the density of defectors 
$x_2$ (blue); note that we slowly increase the parameter $p$ in time at a constant rate, {i.e.}, 
$p$ can be viewed as a time variable. In (a1)-(a2) the minima and maxima of a
moving average are shown whereas (a3) shows the actual time series. The vertical
dashed curve (thin black) in (a1) indicates the theoretically-predicted transition to
oscillations; see (Supplementary Information, Section 3). In (b1) the variances for
$x_{1,2}$ are calculated using a moving window technique up to the gray vertical
lines; note that the scaling of the $V_{1,2}$-axis is $10^{-5}$. Observe that
$x_1$ does not show a clear scaling law while the scaling of $x_2$ can be used for
predicting the transition from steady state to oscillations using classical variance-based warning signs. 
The predicted transition point from extrapolating the increasing variance scaling law \cite{KuehnCT2} is marked as vertical 
dashed line (black) in (a2); note that there is a delay in the Hopf bifurcation point so the predicted critical transition 
to matches, from a practical viewpoint, the data better than the second-order moment closure theory 
\cite{ZschalerTraulsenGross}. In (b2) the
period $T$ of the oscillation is measured and $1/T$ is linearly interpolated to
approximate the period blow-up\cite{Kuznetsov} point (yellow). This is used to predict the transition point (green) from a
periodic to a saddle-type/homoclinic regime; note that this period blow-up is not the saddle-mechanism we focus on
in this paper but another new warning sign we just note as an interesting related result. The predicted 
transition is marked by the dashed vertical line (green) in (a3). Then we also show the logarithmic
distance reduction measured from (a3) in (b3). The ellipses in (b31) indicate the regime
where the decay-scaling for the saddle-approach breaks down. Note that the ellipses are there to guide the
eye. If one would want to give an explicit warning sign, a threshold for $\lambda_u$ has to be specified, 
which is not done in this qualitative example. A detailed quantitative analysis of thresholds is carried 
out for a data set below using ROC analysis. Here we just want to point out the existence of saddles
and the qualitative change in the distance reduction near the saddle, i.e. the parts (a3)
and (b3) illustrate the main ideas for saddle-escape warning signs.}
\end{figure}

We now explore whether the logarithmic scaling discussed above is capable of providing early-warning 
of these outbreaks. We build on full agent-based simulation but our analysis focuses on a pair of 
observables. In particular, we study the density of cooperators, $x_1$, {i.e.}, the proportion of agents whose 
current strategy is cooperation, and the density of links (per agent) that exist between cooperators 
and defectors, $x_2$. Note that per-capita densities of populations are also frequently the only natural
variables available in data, so they are a natural choice to detect critical transitions and warning signs. 

First, let us establish that the outbreaks of defection are indeed triggered by a saddle-escape transition.
Although bifurcation-induced transitions are not the main focus of this paper, we provide a brief description
of the known bifurcation structure here as a parameter is varied and consider the dynamics as a function 
of the rewiring $p$, which measures 
the relative time scale of structural changes of the network to internal changes in the agents strategy. 
This rewiring rate was previously identified as a key parameter of the system (see also Supplementary Information, Section 3). 

Increasing $p$ at a constant rate from a small initial value, we observe two main dynamical changes. First, a stationary
solution turns into stable oscillations, which increase in amplitude and period.
Here we observe the classical increase in variance before a Hopf
bifurcation\cite{Schefferetal,KuehnCT2} (Fig.~\ref{fig:fig2} (b1)), which is a
good predictor for the transition from random fluctuations to small
deterministic oscillations. Second, critical transitions occur at larger
rewiring rates, where the cooperator density $x_1(t)$ drops sharply to lower
values before rising again slowly. These transitions are associated to the
presence of a homoclinic loop in the system\cite{ZschalerTraulsenGross}, which
is attached to the fully cooperative state $x_1^*=x_2^*=1$.

In (Fig.~\ref{fig:fig2} (b2)), we show that the period $T$ of the oscillations may
grow rapidly in finite time, indicating a global bifurcation that gives
rise to the homoclinic loop (see {Ref.}~\cite{Kuznetsov} for further background on 
homoclinic bifurcations). Trajectories close to the homoclinic loop remain near the
saddle point for a long time before making a fast excursion. Locally, near 
$x_1^*=x_2^*=1$, this is precisely the situation of saddle-escapes. The 
logarithmic distances $d_{1,2}$ are shown in (Fig.~\ref{fig:fig2} (b3)). 
As a warning sign to predict the rapid drops in $x_1$ in the range of $p\in[0.93,0.96]$, 
we assume that we know an eventual saddle-instability will happen so 
we just have to look for a change of linear scaling induced by stable directions 
for $d_1$ during the phases when $x_1$ is gradually increasing. These changes can be 
seen for $p\approx 0.944$ and $p\approx 0.957$ as predicted by the theory see (Fig.~\ref{fig:fig2} (b3)).
Note that the complex saddle-escape dynamics occurs in the regime of relatively large
re-wiring rate, when the dynamical process of node update and re-wiring act on similar
time scales, while the bifurcation-induced warning signs worked for low re-wiring in a
quasi-stationary scenario. 

\begin{figure}[htb]\centering
\includegraphics[width=1\textwidth]{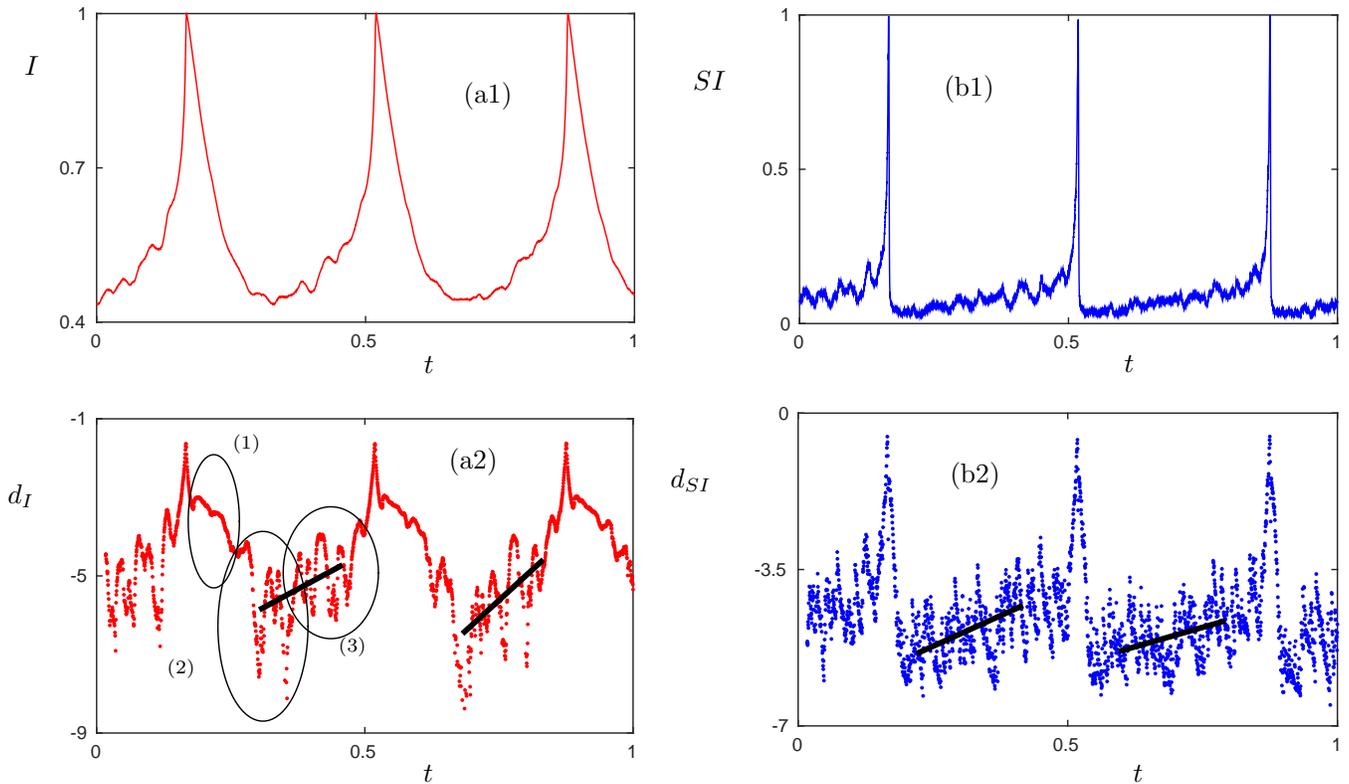}
\caption{\label{fig:fig3}Epidemic outbreaks and prediction in an adaptive SIS
model. (a) Normalized
time series for the infected density $I$ (red) and the susceptible-infected link
density $SI$ (blue). (b) Logarithmic distances for $I$ and $SI$ are shown as
well (see Supplementary Information, Section 2). The linear interpolations 
(black) indicate the expected linear upward trend before a saddle-escape; the slopes of the
four black lines (from left to right) are approximately $7.364$, $12.516$, $5.466$ and $3.461$ respectively. 
The three ellipses in (a2) highlight the three typical regimes between spikes discussed in the
text and are there to guide the eye as in (Fig.2). Parameter values for this figure are $p=0.0058$, 
$r=0.002$ and $w_0=0.6$.}
\end{figure}

As a second example, we consider a susceptible-infectious-susceptible (SIS)
epidemiological model on an adaptive network\cite{GrossDLimaBlasius,GrossKevrekidis}. 
The network consists of susceptible 
(S) and infectious (I) agents. An infection spreads along S-I-links with 
probability $p$ per unit time, infectious agents recover with probability $r$, and 
susceptible agents try to avoid infectious ones by rewiring S-I-links to S-S-links 
with a probability that is proportional to the total number of infectious agents 
(see Supplementary Information, Section 4).

In simulations of this system the number of infectious agents shows
distinguished peaks in time, which can be interpreted as epidemic outbreaks
(Fig.~\ref{fig:fig3}). In (Fig.~\ref{fig:fig3} (b)) the logarithmic distances 
between consecutive points are shown for the density of I-nodes and S-I-links. Despite 
the strong fluctuations away from the peaks, both warning signals show three  
phases after a peak: (1) strong stabilization, (2) plateau- or noise-type behavior and 
(3) a trend towards instability before the next spike (see Fig.~\ref{fig:fig3} (a2)). In fact, 
similar phases can also be observed for the first model in (Fig.~\ref{fig:fig2} (b3)). 

Furthermore, we note that the logarithmic distance increases much earlier for the S-I-links 
than for the I-nodes, so that monitoring the links between infectious and susceptible agents 
provides an earlier warning signal for epidemic outbreaks (Fig.~\ref{fig:fig3}). This is 
in accordance with the intuitive idea that knowledge about the contact dynamics among infectious 
and susceptible agents should allow to predict epidemic outbreaks more easily. 

The two models suggest that there are regions in parameter space where
saddle-escapes play a key role. However, it is also known that epidemic network models can
exhibit bifurcation-induced critical transitions with classical warning signs critical 
transitions\cite{KuehnCT2,OReganDrake}. However, in those cases one usually assumes
that a parameter, {e.g.}~the infection rate, is very slowly varying and this causes the 
critical transition. This directly motivates the question whether there are data sets 
available where epidemics are driven by saddle escapes.

\begin{figure}[htb]\centering
\includegraphics[width=1\textwidth]{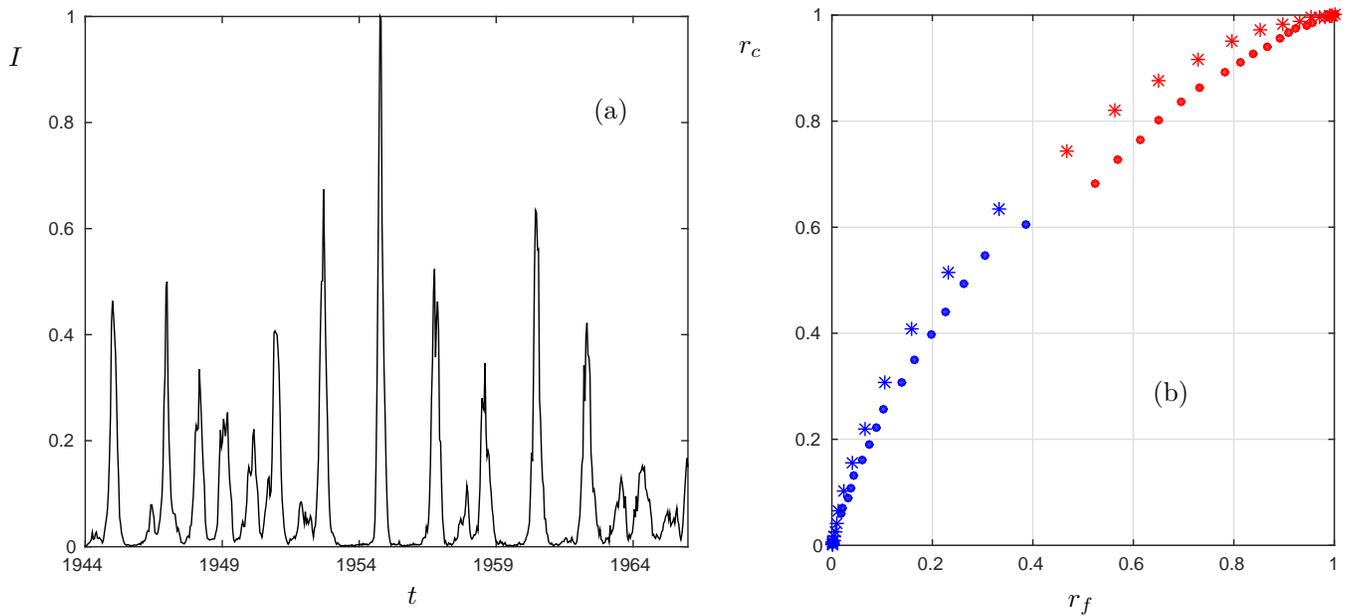}
\caption{\label{fig:fig4}Measles epidemics in the UK between 1944 and 1966. (a)
Typical time series of a city (here: Birmingham) for the infected population
$I$; the series has been normalized by the maximum outbreak. (b) ROC curves for
a five (dots) and ten (stars) data point prediction averaged over all cities. 
The diagonal is shown as well. Blue corresponds to a precursor volume with $\delta>0$ 
and red to a precursor volume with $\delta<0$. A prediction time window of 5 months 
is indicated by 'stars' while a time window of 2.5 months is indicated by 'dots'.}
\end{figure}

Here we focus on repeated measles outbreaks documented biweekly between 1944 and 
1966 in 60 cities in the United Kingdom\cite{BjornstadFinkenstaedtGrenfell,GrenfellBjornstadFinkenstaedt}. 
The time series of the proportion of infected individuals shows long repeated periods of
low disease prevalence interspersed with large, but very short, outbreaks. This strongly indicates 
that a saddle-type mechanism may be at work (Fig.~\ref{fig:fig4} (a)). We use a
receiver-operating-characteristic (ROC) curve\cite{HallerbergKantz,BoettingerHastings} 
to quantify the performance of our warning sign for saddle escapes. We briefly recall, how ROC
curves are calculated. First, one defines a scalar precursory variable $X$ to be computed from
observations before the transitions and considers a threshold $\delta$ such that if $X>\delta$
an alarm is given. Then the two ratios of correct alarms to the total number of actual tipping 
events and false predictions to the total number of non-tipping events are calculated for various
thresholds $\delta$. This provides a quantitative indicator for
the ability of the precursor variable to detect tipping points (see Supplementary Information, 
Section 6, for more background on ROC curves and their interpretation).    

For the ROC curves in our case, we compute a least squares fit of the parameter $\lambda_u=:X$
from equation \eqref{eq:est_u_main} within a time window of $k$ data points as a
precursory variable. We give an alarm for an imminent outbreak at the end of the
time window when $\lambda_u > \delta$ for some threshold $\delta$. This
prediction is considered correct if the disease prevalence exceeds $0.1$ within
the next $5$ data points, which corresponds to an outbreak of at least ten percent
of the maximum outbreak coming up within the next $2.5$ months. In (Fig.~\ref{fig:fig4} (b)) 
we show two ROC curves relating the rate of correct predictions
to the rate of false positives for different threshold values $\delta$. Both curves 
lie above the diagonal, indicating that our method is better than a purely random 
prediction. Moreover, the predictions using
positive threshold values $\delta$ are, on average, better than the ones using
$\delta<0$, as the former produce a ROC curve farther away from the diagonal.
This is a natural result, because $\delta>0$ corresponds to the detection of an
actual instability while $0>\delta\gg -1$ only corresponds to the detection of a
weakly stable direction. Clearly, very short prediction times (small $k$) are 
problematic as they increase the error in $\lambda_u$ leading to false alarms. 
Long prediction times (large $k$) lead to very few correct predictions because
the exponentially stable approach to the saddle strongly dominates on long
time scales. Although the ROC results show that we can potentially improve the 
prediction of epidemic outbreaks, the situation is far from ideal. The ROC 
curve is still quite far from the top-left corner in (Fig.~\ref{fig:fig4} (b)),
which would be perfect prediction. Hence, there is substantial work to be 
carried out to try to increase the practical performance of the warning sign.

Note that, although we demonstrated many factors which strongly 
indicate a saddle-escape mechanism in measles data, it is unlikely that a test 
exists which can guarantee detailed knowledge of the underlying dynamical mechanisms 
based on just a relatively short uni-variate time series of the epidemic. However, 
there is evidence from several epidemic models, which do display saddle 
states\cite{DerrickVandenDriessche,VandenDriesscheWatmough}. It then remains as
an open problem to link our warning sign analysis of measles data to particular
models of measles\cite{RohaniKeelingGrenfell,FinkenstaedtGrenfell}, which we 
leave as a challenge for future work. Furthermore, we remark that 
although the indicator we have used seems reasonably efficient, its robustness will 
strongly depend on the noise level in the system. 

As a general comment, we note that the exact mechanism by which a critical transition occurs 
may well lie in the eye of the beholder. A given critical transition observed in nature 
may appear as a bifurcation-induced transition in one model and as a saddle-escape in 
a different model describing the same phenomenon with different variables.
  
In summary, we have identified a new type of critical transitions that could be relevant for 
real-world complex many-variable saddle-type systems. We proposed a simple, intuitive early warning signal 
for these transitions, and demonstrated the application of this warning signal in two adaptive network models 
and real-world data. We believe that this warning signal will be useful for detecting saddle-escape transitions 
in network models and data sets, for instance for ecological and socio-economic systems. Because the underlying 
mechanism considered here differs from the ones typically studied in critical transitions, the proposed early 
warning sign is complementary to existing approaches. 

It is crucial to point out that significant
additional research is necessary to make warning signs, for saddle-escape as well as the classical
bifurcation-induced tipping scenarios, more applicable for many real-world applications. The first main
issue is to connect results to a more detailed statistical analysis, which is currently work in 
progress by several groups\cite{BoettingerHastings}.
A second challenge is to link qualitative modelling results better to quantitative warning signs. Indeed,
both necessarily depend on each other for saddle- and bifurcation-mechanisms. In both cases the predictions
are tremendously improved if one knows \textit{a priori} that the current dynamical evolution
of the system may be towards a destabilizing tipping scenario and one just does not know \textit{where}
the system tips, on what \textit{time horizon} it happens, and which precise \textit{mechanism} occurs. 
A third challenge is to keep new potential applications in perspective and there has
been significant recent progress in this direction in realistic systems for bifurcation-induced 
transitions, for example in ecology\cite{Veraartetal}.
Various models in neuroscience\cite{Izhikevich} and in ecology\cite{Hastings7} suggest that saddle points can 
occur, which are bound to be preceeded by the generic logarithmic distance reduction we use as a 
warning sign. In an ecological context a natural future test case could be 
bio-invasions\cite{Bright2,PhilippsBrownWebbShine}.\medskip 

\textbf{Acknowledgements:} {C.K.}~would like to thank the Austrian Academy of Sciences ({\"{O}AW}) for 
support via an APART fellowship as well as the European Commission (EC/REA) for support by a Marie-Curie 
International Re-integration Grant. The majority of this work was carried out while all three authors worked 
at the Max Planck Institute for the Physics of Complex Systems (MPI-PKS); hence we would like to acknowledge 
it here for hospitality and financial support.

\part*{Supplementary Information}
 \section{Metastability Near Saddle Points}
 
 \subsection{Abundance of Saddles in Complex Systems}
 \label{ssec:abundance}
 
 Our main assumptions on complex networks is that steady states are
 ``frequently'' saddle points. In this section, we provide a mathematical
 justification for this assumption. Suppose we are given a large system of ODEs
 \be
 \label{eq:ap_main_eq}
 x'=f(x)
 \ee
 for $x\in\R^n$ and $n\gg 1$. Assume $x^*$ is a steady state (or equilibrium
 point, $f(x^*)=0$) for \eqref{eq:ap_main_eq}. Denote the linearization (or Jacobian) $Df(x^*)$
 by $A_n\in\R^{n\times n}$ which generically provides the local stability of
 $x^*$ using the Hartman-Grobman Theorem\cite{GH}. Eigenvalues
 of $A_n$ with negative/positive real parts correspond to stable/unstable
 eigendirections. Under the hypothesis that the system is complex, heterogeneous and its size 
 is large it is reasonable to assume\cite{May1} that $A_n$ is a random matrix with independent
 identically distributed (iid) entries given by a complex random variable $z$
 with mean zero and variance $\sigma^2$. Let $\{\lambda_i\}_{i=1}^n$ denote the
 eigenvalues of $A_n$ and define the empirical spectral distribution $\mu_n$ of
 $A_n$ by
 \benn
 \mu_n(s,t):=\frac1n\#\{k\leq n:\text{Re}(\lambda_k)\leq s\text{ and
 }\text{Im}(\lambda_k)\leq t \}.
 \eenn
 where $\#$ denotes the cardinality of a set. Recently it has been proven 
 (see\cite{TaoVu} and references therein) that, under suitable boundedness assumptions 
 on the moments of $z$, the circular law conjecture holds which states that $\mu_n$ converges 
 to the uniform distribution over the unit disk
 \benn
 \mu_\I(s,t):=\frac1\pi \text{mes}(\{w\in \C:|w|\leq 1, \text{Re}(w)\leq s\text{
 and }\text{Im}(w)\leq t \})
 \eenn
 where '$\text{mes}$' denotes Lebesgue measure. Hence, asymptotically as $n\ra \I$,
 if $\lambda$ is an eigenvalue of $A_n$ then
 \benn
 \P(\text{Re}(\lambda)<0)=\frac12=\P(\text{Re}(\lambda)>0).
 \eenn
 This implies that $x^*$ is stable with probability $(1/2)^n$, completely
 unstable with probability $(1/2)^n$ and a saddle point with probability
 $1-(1/2)^{n-1}$. Therefore, the probability that $x^*$ is a saddle point tends
 to one as $n\ra \I$. 
 
 It is extremely important to note that the argument here
 is based upon certain mathematical assumptions to make it rigorous. However, it
 is strongly expected that if we \emph{weaken} the assumptions in various ways, 
 we still find saddle points frequently in high-dimensional systems. In fact, let 
 us point out that the idea to characterize instability in large-scale
 systems using random matrix theory is well-known\cite{GardnerAshby} but is still
 a topic of very recent interest\cite{WainribTouboul}. However,
 previously one only had the semi-circular law available that required the
 symmetry of $A_n$ or one had to rely on structured matrices, for example certain
 types of food webs\cite{May1}. These assumptions are usually too strong for
 complex dynamical networks, which can be highly heterogeneous and yield
 unstructured, non-symmetric ODEs. This makes the recent progress on proving the
 full circular law conjecture important in our context. As discussed above, one 
 expects that the results can be generalized even further to include even larger
 classes of complex systems. Furthermore, note carefully that even if $n$ is small, 
one may still have saddles, which may be relevant for the dynamics, {i.e.}, the 
assumptions we make are \textit{sufficient} to prove saddle existence with high 
probability but the assumptions may not be \textit{necessary} to find saddles at 
all. 
 
 \subsection{Residence Times}
 \label{ssec:residence}
 
 Another main point of our argument is that the systems can spend a much longer
 time near saddle points than away from them. This leads to metastable behavior
 near saddle points. This can be illustrated with the simplest two-dimensional
 case given by the ODEs
 \be
 \label{eq:plain_saddle}
 \begin{array}{lcl}
 x_1'&=&\lambda_s x_1,\\
 x_2'&=&\lambda_u x_2,\\
 \end{array}
 \ee
 where $\lambda_s<0<\lambda_u$. Note carefully that it is justified to reduce the 
 dimension of the system \emph{after} the large dimensionality of the system has led
 to saddle points; the mathematically rigorous reduction just follows from center 
 manifold theory \cite{GH}. The system \eqref{eq:plain_saddle} decouples with
 solution
 \benn
 x_1(t)=x_1(0)e^{\lambda_s t} \qquad\text{and}\qquad x_2(t)=x_2(0)e^{\lambda_u
 t}.
 \eenn
 Suppose we start with some $x_1(0)=\kappa>0$ and want to reach a small
 neighborhood of the origin with $x_1(T)=\delta\ll \kappa$. This takes a time
 $T=\lambda_s^{-1} \ln(\delta/\kappa)$. Viewing $T$ as a function of $\kappa$
 shows that the time increases logarithmically. Therefore, a trajectory spends a
 much longer time near the equilibrium in comparison to the approach towards the
 equilibrium. Similarly, we can require a trajectory to start in a small
 neighborhood of the saddle with $x_2(0)=\delta>0$ and end at $x_2(T)=\kappa>0$. Then
 $T=\lambda_u^{-1}\ln(\kappa/\delta)$ and the same arguments apply to show that
 the initial time spend near the equilibrium is much longer than the escape time.
 Although we have only worked with a linear system \eqref{eq:plain_saddle}, similar
 conclusions apply for the nonlinear case as long the passage near the hyperbolic saddle 
 occurs sufficiently close to its stable and unstable manifolds\cite{GH}.
 
 \section{Saddle Point Warning Signs}
 \label{sec:saddle_warning}
 
 \subsection{Basics}
 
 Locally near a hyperbolic saddle point, which we can assume without loss of
 generality to be at $x=(0,0,\ldots,0)=:0$, we can work with the linearization so
 that
 \be
 \label{eq:linear_sp}
 x'=Ax
 \ee
 for some matrix $A\in\R^{n\times n}$ with eigenvalues $\lambda_i\in \C$ for
 $i\in\{1,2,\ldots,n\}$ and associated eigenvectors $v_i$. We are going to assume
 that the eigenvalues are distinct which is generic within the space of matrices.
 Standard linear algebra gives a coordinate transformation $P:\R^n\ra \R^n$,
 $x=Py$, so that
 \benn
 P^{-1}AP=B=\left(
 \begin{array}{ccc}
 B_1 & & \\
 & \ddots & \\
 & & B_k \\
 \end{array}
 \right)
 \eenn
 where the $k$ matrices $B_1,\ldots,B_k$ are the usual Jordan blocks 
 and $P$ maps the standard basis vectors to the basis
 $\{v_i\}$. Since the eigenvalues are distinct we have $B_j\in\R$ or
 $B_j\in\R^{2\times 2}$. It is straightforward to observe that the escape near
 saddles is governed by the weakest stable and the strongest unstable directions.
 More precisely, we will only consider at most four eigenvalues
 $\lambda_{s},\overline{\lambda_s},\lambda_u,\overline{\lambda_u}$ where overbar
 denotes complex conjugation so that
 \be
 \label{eq:eig_assume}
 \begin{array}{ll}
 0>\text{Re}(\lambda_s)>\text{Re}(\lambda_k), &\qquad\text{for all $k\neq s$ such
 that $0>\text{Re}(\lambda_k)$},\\
 0<\text{Re}(\lambda_k)<\text{Re}(\lambda_u), &\qquad\text{for all $k\neq u$ such
 that $0<\text{Re}(\lambda_k)$}.\\
 \end{array}
 \ee
 It is extremely important to highlight again the logic in the previous 
 derivations: First, we start with a large-dimensional system, where it can
 be shown that saddle points are frequent. For each hyperbolic saddle point,
 there are many eigenvalues. However, for the dynamical approach or departure
 of the saddle point, the dynamics is locally governed by the weakest stable 
 and strongest unstable directions, {i.e.}, stronger stable directions damp out 
 very quickly, while weak unstable directions are generically dominated by the
 strongest unstable mode. Hence, we may develop a local theory for high-dimensional
 hyperbolic saddles by focusing on the leading directions in the stable and 
 unstable manifolds, which are generically low-dimensional.
 
 \subsection{The Planar Saddle}
 \label{ssec:n2}
 
 We start with the case $n=2$ and $\lambda_{s,u}\in\R$. Setting $x=Py$ gives
 $y'=P^{-1}APy=By$ with solution $y(t)=y(0)e^{tB}$ or $x(t)=Py(t)$ so that
 \benn
 x(t)=y_1(0)e^{\lambda_s t}v_1+y_2(0)e^{\lambda_u t}v_2.
 \eenn
 for vectors $v_{1,2}$ that can be calculated explicitly. Since we want to
 approach the saddle point and stay near it for some significant amount of time
 we must have that $|y_2(0)|\neq 0$ is small so that a solution starts close to
 the stable manifold $W^s(0)=\{\rho v_1:\rho\in\R\}$. Let $\|\cdot\|$ denote the 
 usual Euclidean norm. Observe that the term
 $\|y_1(0)e^{\lambda_s t} v_1\|\ra 0$ as $t\ra \I$ and, if $y_2(0)\neq0$,
 $\|y_2(0)e^{\lambda_u t} v_2\|\ra \I$ as $t\ra \I$. Hence, initially $\|x(t)\|$
 decreases exponentially until a unique minimum and then $\|x(t)\|$ increases
 exponentially. For times $t_j>0$ such that $y_2(0)e^{\lambda_u t_j}$ is small it
 follows for some $t_2>t_1>0$ that
 \benn
 \|x(t_2)-x(t_1)\|\approx |y_1(0)||e^{\lambda_s t_2}-e^{\lambda_s
 t_1}|\|v_1\|=|y_1(0)|\|v_1\||e^{\lambda_s t_1}(e^{\lambda_s (t_2-t_1)}-1)|
 \eenn
 If $t_2\gg t_1$ then $e^{\lambda_s (t_2-t_1)}\approx 0$ so that
 \be
 \label{eq:est_s}
 \ln \|x(t_2)-x(t_1)\|\approx \lambda_s t_1+k_1
 \ee
 where $k_1=\ln(|y_1(0)|\|v_1\|)$ is a constant that will not be of relevance
 here. Observe that \eqref{eq:est_s} allows us to estimate $\lambda_s$ from data.
 Then we can consider it a warning sign when the logarithm of the distance
 between points starts to deviate from the linear fit \eqref{eq:est_s}. In the
 regime where $|y_1(0)|e^{\lambda_s t}\|v_1\|$ is small a similar procedure
 allows us to estimate the strongest unstable eigenvalues since then we find
 \be
 \label{eq:est_u}
 \ln \|x(t_2)-x(t_1)\|\approx \lambda_u t_2+k_2
 \ee
 where $k_2=\ln(|y_2(0)|\|v_2\|)$. From the knowledge of $\lambda_u$ we can
 predict how rapidly $\|x(t)\|$ is expected to grow. In practice, we can estimate
 the eigenvalues $\lambda_{s,u}$ from a uni-variate coordinate time series $x_i(t)$ by
 looking at a fixed time point $T$ and a set of $K$ previous times
 $t_1<t_2<\cdots<T$ to compute
 \be
 \label{eq:dlog}
 d_i(T):=\frac1K \sum_{k=1}^K\ln |x_i(T)-x_i(t_k)|.
 \ee
 Computing $d_i(T)$ for different times $T$ gives that in different regimes
 (stable/unstable) we have $d_i(T)\sim \lambda_{u,s}T+K_2$ for some constant $K_2$.
 
 \subsection{Complex Eigenvalues}
 
 For the case $n=3$ we will again assume that \eqref{eq:eig_assume} holds and
 that the complex conjugate eigenvalue pair has negative real part i.e. we
 consider $\lambda_s=a_s+ib_s$, $\overline{\lambda_s}=a_s-ib_s$, $\lambda_u$ with
 associated real eigenvectors $v_{1,2,3}$. With $x(t)=Py(t)$ the general solution
 is
 \benn
 y(t)=\left(
 \begin{array}{ccc}
 e^{a_st}[y_1(0) \cos(b_st)+y_2(0)\sin(b_st)]\\
 e^{a_st}[y_2(0) \cos(b_st)-y_1(0)\sin(b_st)]\\
 y_3(0)e^{\lambda_u t}
 \end{array}
 \right).
 \eenn
 Writing the solution for $x(t)$ in the basis of the eigenvectors $v_i$ gives
 \beann
 x(t)&=&e^{a_st}[y_1(0) \cos(b_st)+y_2(0)\sin(b_st)]v_1\\
 &&+e^{a_st}[y_2(0)
 \cos(b_st)-y_1(0)\sin(b_st)]v_2+y_3(0)e^{\lambda_u t}v_3.
 \eeann
 As before, we are going to distinguish two regimes in the time domain, starting
 with the assumption that $\|y_3(0)e^{\lambda_u t}v_3\|$ is small which yields
 exponentially decaying oscillations in time series for each coordinate $x_i$.
 Let $T$ be the time between successive maxima or minima then $b_s=2\pi/T$.
 Furthermore, if $t_2>t_1>0$ as previously and $t_2-t_1=T$ then
 \benn
 \|x(t_2)-x(t_1)\|\approx k_3 |e^{a_st_2}-e^{a_st_1}|
 \eenn
 for a positive constant $k_3$. The last equation can then be used to estimate
 $a_s$ as shown in Section \ref{ssec:n2}. The case of two complex conjugate
 eigenvalue pairs is similar and will not be discussed here.
 
 \subsection{Noisy Saddles}
 
 An important question is to consider the influence of noise as natural systems,
 and in particular the measurement of natural systems, are often well-described
 by an underlying deterministic system with additional random fluctuations.
 Consider the standard one-dimensional Ornstein-Uhlenbeck (OU) $x=x(t)$ stochastic process 
 generated by the stochastic differential equation (SDE)
 \be
 \label{eq:OU}
 \text{d}x=\frac12ax~\text{d}t+\sigma ~\text{d}W
 \ee
 where $W=W(t)$ is a standard 1-dimensional Brownian motion and the equation is interpreted
 in the It\^o-sense. It is well-known\cite{Gardiner} 
 that the solution to \eqref{eq:OU} and the resulting variance
 $V(t)=\text{Var}(x(t))$ can be calculated
 \benn
 V(t)=\frac{\sigma^2}{a}\left(e^{at}-1\right).
 \eenn
 For $a<0$ it follows that $V(t)\ra -\sigma^2/a$ as $t\ra \I$ and for $a>0$ one
 gets $V_t\ra \I$ as $t\ra \I$. More precisely,
 \be
 \label{eq:scale_Var}
 \ln(V(t))=\ln \sigma^2 -\ln a+\ln\left(e^{at}-1\right)\sim 2\ln \sigma -\ln a+a t
 \qquad \text{as $t\ra \I$.}
 \ee
 Generalizing \eqref{eq:OU} to the simplest possible saddle point yields
 \be
 \label{eq:OU1}
 \text{d}x=Ax~\text{d}t+\sigma~ \text{d}W
 \ee
 where $W=W(t)$ now denotes a standard 2-dimensional Brownian motion and $A$ has two
 eigenvalues $\lambda_s<0<\lambda_u$. Then the same conclusion as before apply
 since the entries of the covariance matrix $C(t)=\text{Cov}(x(t))$ are generically
 linear combinations of two decoupled OU-processes, one stable with
 asymptotically constant variance for $a=\lambda_s$ and one with diverging
 variance for $a=\lambda_u$. Hence one could also attempt to use the scaling
 \eqref{eq:scale_Var} to get an estimate for $\lambda_u$ by considering a moving
 window analysis of the logarithm for the variance. This could be of particular
 interest in case the more straightforward logarithmic distance reduction method
 does not work. Most likely this will be the case only for very particular
 intermediate noise strengths. For small noise the estimator based on distances
 works quite well as shown in the main manuscript. However, if the noise is too
 large one never reaches a neighborhood of the saddle point with
 high-probability so that predictions become impossible anyway i.e.~the events 
 acquire a purely noise-induced character.
 
 \subsection{An Example from Epidemics} 
 
Saddle points also appear frequently in many applications. Here we briefly illustrate 
the dynamics near saddles for a compartmental epidemic 
model\cite{DerrickVandenDriessche,LiuLevinIwasa}. The basic model is given by
\be
\label{eq:epmodLLI}
\begin{array}{lcl}
S'&=& −\phi I^2 S/T^2 − d S + \rho R + bT,\\
I'&=& \phi I^2 S/T^2 − (d + \gamma)I,\\
R'&=& \gamma I − (d + \rho)R,\\
\end{array}
\ee
where $S$, $I$, $R$ are the number of susceptible, infected and recovered 
individuals in the population (with $T := S + I + R$), $d$ and $b$ are per 
capita death and birth rates, $\gamma$ is the per capita recovery rate, and 
$\rho$ is the per capita loss of immunity constant. $\phi$ is the main 
bifurcation parameter of the model and controls the interaction strength 
for the nonlinear incidence function $I^2 S$. Assuming $d = b$ to
keep the total population constant, introducing the new variables
\benn
s := S/T, \qquad i := I/T, \qquad r = R/T,
\eenn
and using the constraint $1 = s + i + r$, one arrives at a two-dimensional 
ODE system\cite{DerrickVandenDriessche}
\be
\label{eq:DDep}
\begin{array}{lcl}
i' &=& −i[\phi i(1 − i − r) − (b + \gamma)],\\
r' &=& \gamma i − (b + \rho)r.\\
\end{array}
\ee

\begin{figure}[htb]
\centering
\includegraphics[width=1\textwidth]{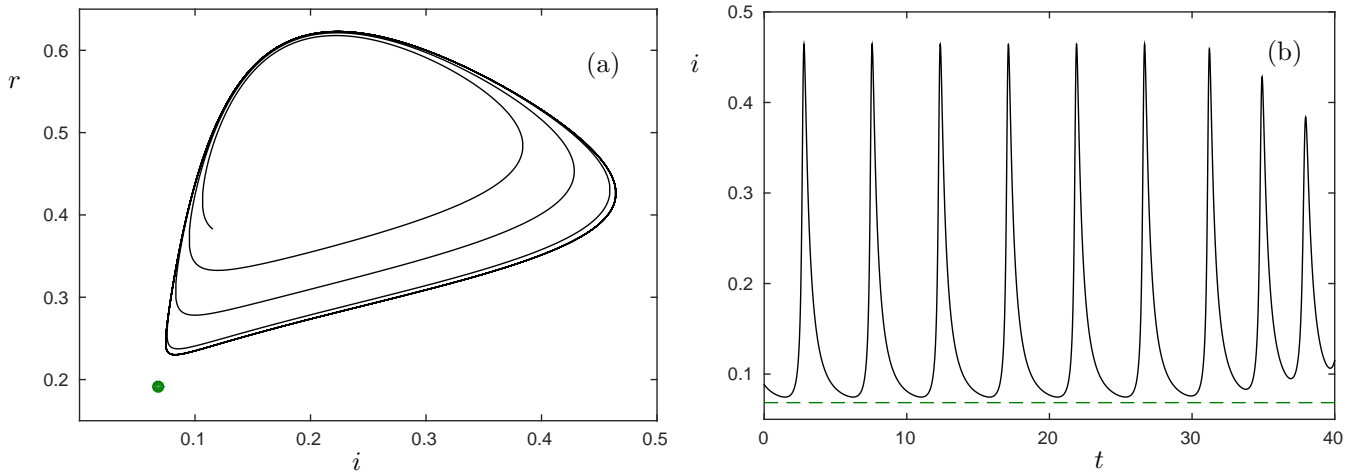}
\caption{\label{fig:fig5}Numerical simulation for the compartmental epidemic 
model \eqref{eq:DDep}. (a) Phase space plot with saddle point (green dot) and 
a trajectory segment (black curve) showing multiple epidemic outbreaks with 
passages near a saddle. (b) Time series for the infected population density $i$ 
corresponding to the trajectory from (a). The $i$-value of the saddle
point is indicated by a dashed green line. The long passages near the saddle between
larger epidemic ourbreaks are clearly visible on this time scale.}
\end{figure}

The system \eqref{eq:DDep} has a number of different dynamical regimes 
depending upon the parameter values. Figure \ref{fig:fig5} shows a simulation 
of the dynamics for parameter values $b = 1$, $\rho = 0.132051$, $\phi = 81.88$, 
$\gamma = 1.4$. In the $(i,r)$-phase space plot in Figure \ref{fig:fig5}(a) a saddle point 
$(i, r) = (i^∗ , r^∗ )$ has been marked as a dot. The trajectory approaches a 
neighbourhood of the saddle several times, where it only evolves slowly as discussed 
above. These long periods near the saddle are then interspersed with several 
short periods consisting of large epidemic outbreaks. The dynamics in this parameter 
regime is transient and will settle after a very long time to a stable sink equilibrium. However, 
before this occurs, the saddle dynamics plays the main role. Figure 5(b) shows a time series 
of the infected population density $i$ corresponding to the trajectory from Figure \ref{fig:fig5}(a).

It is important to note that the observed effect can also occur for the case when the saddle
point lies precisely on the zero line. Indeed, if we consider a coordinate change $\tilde{i}=i-i^*$, 
then the saddle point lies on the zero line $\{i = 0\}$ and we observe the same saddle escape phenomenon
as above.

 \section{Cooperation Games on Networks}
 \label{sec:coop_nets}
 
 In this section we give a more detailed technical description of the snowdrift
 game network. The evolutionary game is
 defined between agents (nodes) that interact on an adaptive network via the
 links between them\cite{ZschalerTraulsenGross}. The number of nodes $N$ and
 number of undirected links $K$ is fixed. However, as described below, the
 adjacency matrix $A=(a_{ij})$ may change in time and interacts with the
 dynamics. This makes the system and adaptive, or co-evolutionary, network. 
An agent $i$ interacts with an agent $j$ at a given time step if there
 is a link between $i$ and $j$. The interaction takes place via a game between
 the two nodes. In this game, an agent can have two possible strategies
 $\sigma_i$, cooperation $C$ or defection $D$, which are the two dynamical states
 of the nodes. The payoff agent $i$ receives from agent $j$ via an interaction is
 modeled via the snowdrift game\cite{DoebliHauert} with interaction matrix
 \benn
 M=\left(\begin{array}{cc}b-c/2 & b-c \\ b & 0\\\end{array}\right)
 \eenn
 where $c$ represents the cost of cooperation and $b$ the benefit. $M_{11}$
 represents cooperation of both agents, $M_{22}$ defection of both agents and the
 off-diagonal entries correspond to the mixed cases where one agent tries to
 cooperate but the other agent defects. The total payoff $\pi_i$ for node $i$ is
 \benn
 \pi_i=\sum_{j:a_{ij}=1} M_{ij}.
 \eenn
 The network is made adaptive by a probabilistic rule. After the game has been
 played, choose a link at random. With probability $p$ re-wire this link and with
 probability $1-p$ one of the linked agents adopts the other agent's strategy. The
 two events of re-wiring and adaptation have to specified in more detail. Define
 the performance $\phi(\sigma)$ of a strategy $\sigma\in\{C,D\}$ as
 \benn
 \phi(\sigma):=\frac{1}{n_\sigma N}\sum_{i:\sigma_i=\sigma}\pi_i
 \eenn
 where $n_\sigma$ is the fraction of agents using strategy $\sigma$. If the
 strategy adoption event takes place agent $j$ adopts the strategy of agent $i$
 with probability
 \benn
 f_\beta(i,j)=\left(1+e^{-\beta[\phi(\sigma_i)-\phi(\sigma_j)]}\right)^{-1}
 \eenn
 and $i$ adopts $j$'s strategy with probability $f_\beta(j,i)=1-f_\beta(i,j)$.
 For a re-wiring event of a link between $i$ and $j$, delete it and select a
 random node $k$. Then the link between $k$ and $i$ is generated with probability
 $f_\alpha(i,j)$ and between $k$ and $j$ with probability $f_\alpha(j,i)$.
 
 The main dynamical variables we are interested in are the fraction of
 cooperators and defectors $n_C$ and $n_D$, as well as the link densities
 $l_{CC}$, $l_{CD}$ and $l_{DD}$. Since the number of nodes and links is constant
 it suffices to restrict attention to a single node density and two link
 densities. Consider the parameter set
 \benn
 \alpha=30,\quad \beta=0.1,\quad b=1,\quad c=0.8, \quad N=50000, \quad K=500000
 \eenn
 where the re-wiring rate $p$ is the primary bifurcation parameter that is varied
 between $p=0$ and $p=1$. There are three main dynamical regimes, for small $p$
 the densities $n_{CD}$ remain almost constant and there are just finite-size effect
stochastic fluctuations. Note that for small $p$, the network topology is very close 
to being static. Increasing $p$ gives rise to a supercritical Hopf bifurcation to 
oscillations in a system, where the population and link densities between different 
types of agents are taken into account\cite{ZschalerTraulsenGross}, {i.e.}~the 
linearization at the near-homogeneous steady state has a pair of complex conjugate 
eigenvalues, which cross the imaginary axis at nonzero speed upon variation of $p$. 
This leads to small-scale deterministic oscillations, which grow in amplitude upon 
increasing $p$ further. For large $p$ the periodic dynamics approaches a near-homoclinic 
orbit with saddle-type escape dynamics and long periods of high cooperation values with 
$n_C$ near $1$. More precisely, the period of the oscillations increases and long times
are spend near a saddle steady state and eventually the periodic orbit limits onto a 
homoclinic orbit, which means trajectories come extremely close to the saddle steady
state. Note carefully that high values of $p$ mean that the network topology can change
very quickly, {i.e.}~quickly in comparison to the changes of the dynamical states of 
the agents. This leads to a highly heterogeneous complex system, which can drastically
change its entire topology and dynamical state; for a more detailed description of the 
dynamics we refer to\cite{ZschalerTraulsenGross}.
 
 \section{Epidemics on Networks}
 \label{ap:epidemics}
 
 In this section we provide a more detailed overview of the epidemiological network
 model discussed in the main text. The total number of nodes $N$
 and links $L$ is assumed to be fixed. Nodes can be in either in a susceptible
 (S) or infected (I) state. Links between nodes represent potential transmission
 routes of the disease. Loops and double-links are not allowed. At each time step
 an infected node recovers with a probability (or recovery rate) $r$ into a
 susceptible node. For every $SI$-link the disease spreads with probability $p$
 so that the $S$ node becomes an $I$ node upon infection. This basic dynamics
 just represents the standard susceptible-infected-susceptible 
 model\cite{BrauervandenDriesscheWu}. In addition, susceptibles can try to avoid
 contact with infected and this is modeled via a probability $w$ of re-wiring an
 $SI$-link. In this case, the susceptible node $S$ cuts its link to a node
 $I$ and establishes a new link to another susceptible node.
 
 As before, we are mainly interested in the node and link densities and their
 trajectories. It is natural to view the parameters $(r,p,w)$ as bifurcation
 parameters. We briefly describe the dynamics that have been found 
 in\cite{GrossDLimaBlasius}. It is observed that cluster formation and degree
 correlation depend on the re-wiring $r$, {e.g.}~higher re-wiring rates lead to
 higher degree correlation. Furthermore, the main bifurcation point upon varying
 $w$ is the epidemic threshold corresponding to a transcritical bifurcation. Hopf bifurcations, 
 saddle-node bifurcations, oscillations and hysteresis can occur.
 
 The re-wiring mechanism can be refined by introducing awareness of susceptibles
 to the disease. Let $\rho=i/N\in[0,1]$ denote the infected fraction of the
 population. Now define $w=w_0\rho$ where $w_0$ is a fixed constant. In addition
 to the bifurcation phenomena observed previously, a homoclinic bifurcation is
 found upon varying $p$. The large-amplitude oscillations that occur near the
 homoclinic bifurcation are of interest for our study of saddle escapes in the main text.
 
 The parameter values used for full network simulation are $p=0.0058$, $r=0.002$
 and $w_0=0.6$ with an initial susceptible density $s_{t=0}=0.98$. The total
 number of nodes was fixed to $N=10^5$ and the total number of links to $L=10^6$. These yield
 oscillations and their analysis using saddle-type escape dynamics gives the
 results shown in (Fig.~\ref{fig:fig3}).
 
 \section{Measles Data}
 
 The data set we use for testing our methods are measles epidemic data recorded
 in 60 UK cities from 1944 to 1966; the data has been downloaded from the 
 website\cite{GrenfellWeb,GrenfellWeb1}. A detailed description and analysis of this data set 
 is given in the two papers\cite{GrenfellBjornstadFinkenstaedt,BjornstadFinkenstaedtGrenfell}. For each
 city the total number of cases has been reported biweekly. Depending on the size
 of the city, measles is expected to occur in endemic cycles (large cities) or in
 recurrent epidemics with local extinction. It is important to observe that
 either of those two phenomena depends on the fact that the virus is somewhere in
 the total host population.
 Therefore, a low infected density is generically expected to lie very close to
 the stable manifold of a saddle point when considered in a sufficiently large
 phase space.
 
 An interesting aspect of the modeling of 
 {Grenfell et al.}\cite{GrenfellBjornstadFinkenstaedt,BjornstadFinkenstaedtGrenfell} is that,
 although their model is stochastic, they recognize that during the initial phase
 and throughout the epidemic, the underlying dynamical system seems to be behave
 deterministically. This is precisely the same behavior one can observe from a
 model such as the epidemiological adaptive-network model described in the last section.
 
 \section{ROC curves}

 Here, we briefly review the main idea of ROC (receiver operating characteristic), also called 
 the ROC curve, and give the formal definitions for the general case. 
 Denote the points in a given time series by $I_j:=I(t_j)\in\R$ for $j=1,2,\ldots$
 and let $Y_{m}$ be a binary random variable with
\benn
Y_m:=\left\{
\begin{array}{ll}
1 & \text{a tipping/event occured at time $m$}\\
0 & \text{no tipping/event occured at time $m$},
\end{array}
\right.
\eenn
{i.e.}~$Y_m$ just records whether a tipping point occured at time $m$ or not. $Y_m$ is
taken as a random variable since we do not a priori when events occur. Consider some 
subset of previous observations 
$I_{k_1,k_2}:=(I_{m-k_1},I_{m-k_1-1},\ldots,I_{m-k_2})\in\R^{k_1-k_2+1}$
with $0<k_2<k_1$, where $k_1-k_2+1$ is also referred to as the sliding window length or 
observational window length.
 
Next, one defines a precursory variable $X_m:=\text{pre}(I_{k_1,k_2})$,
where $\text{pre}:\R^{k_1-k_2+1}\ra \R$ is a mapping, which computes out of the 
observations a scalar-valued precursor. Obviously the variable $X_m$ can depend upon 
the choice of $k_1, k_2$ and on further parameters (this problem is currently being studied
by the first author and several colleagues for the case of B-tipping). We give an alarm when $X_m>\delta$ for the event
some $\delta\in\R$. The precursory variable enables us to calculate the rate of
correct predictions as well as the rate of false positives
 \benn
 r_c=\frac{\#\text{correct
 predictions}}{\#\text{events/outbreaks}}\quad\text{and}\quad
 r_f=\frac{\#\text{false positives}}{\#\text{non-events}}.
 \eenn
Both rates will obviously depend upon $k_1$, $k_2$, $\delta$ and the dependence will be 
upon $\delta$ only if the window length is fixed. One may also write $r_c$ and $r_f$ by
using aposterior probability density functions as\cite{HallerbergKantz}
\benn
r_c=\int_{\{X_m>\delta\}}\P(X_m|Y_m=1),\qquad r_f=\int_{\{X_m>\delta\}}\P(X_m|Y_m=0).
\eenn
The ROC-curve is a plot of the rates in the $(r_f,r_c)$-plane for different values of 
the threshold $\delta$; see (Fig.~\ref{fig:fig4}) for an example. There are several
important standard observations about ROC curves. Perfect prediction occurs if
no false positive and all true events are detected. This implies that the ROC
curve should consist just of the point $(r_f,r_c)=(0,1)$. The point $(0,0)$ in the 
lower-left corner of an ROC curve represents a value of $\delta$ that is so high that
no alarm is given at any point while the point $(1,1)$ at the upper right corner 
represents when alarms are given at every time step. The diagonal connecting 
$(0,0)$ and $(1,1)$ is precisely, where true positive rate equals
the false positive rate, which is equivalent to making random
guesses. A precursor with performance better than random guesses 
corresponds to a point in the upper triangle with $r_c > r_f$.


\begin{thebibliography}{10}
\expandafter\ifx\csname url\endcsname\relax
  \def\url#1{\texttt{#1}}\fi
\expandafter\ifx\csname urlprefix\endcsname\relax\def\urlprefix{URL }\fi
\providecommand{\bibinfo}[2]{#2}
\providecommand{\eprint}[2][]{\url{#2}}

\bibitem{Schefferetal}
\bibinfo{author}{Scheffer, M.} \emph{et~al.}
\newblock \bibinfo{title}{Early-warning signals for critical transitions}.
\newblock \emph{\bibinfo{journal}{Nature}} \textbf{\bibinfo{volume}{461}},
  \bibinfo{pages}{53--59} (\bibinfo{year}{2009}).

\bibitem{Lentonetal}
\bibinfo{author}{Lenton, T.} \emph{et~al.}
\newblock \bibinfo{title}{Tipping elements in the {Earth's} climate system}.
\newblock \emph{\bibinfo{journal}{Proc. Natl. Acad. Sci. USA}}
  \textbf{\bibinfo{volume}{105}}, \bibinfo{pages}{1786--1793}
  (\bibinfo{year}{2008}).

\bibitem{KuehnCT1}
\bibinfo{author}{Kuehn, C.}
\newblock \bibinfo{title}{A mathematical framework for critical transitions:
  bifurcations, fast-slow systems and stochastic dynamics}.
\newblock \emph{\bibinfo{journal}{Phys. D}} \textbf{\bibinfo{volume}{240}},
  \bibinfo{pages}{1020--1035} (\bibinfo{year}{2011}).

\bibitem{GH}
\bibinfo{author}{Guckenheimer, J.} \& \bibinfo{author}{Holmes, P.}
\newblock \emph{\bibinfo{title}{Nonlinear Oscillations, Dynamical Systems, and
  Bifurcations of Vector Fields}} (\bibinfo{publisher}{Springer},
  \bibinfo{address}{New York, NY}, \bibinfo{year}{1983}).

\bibitem{Kuznetsov}
\bibinfo{author}{Kuznetsov, Y.}
\newblock \emph{\bibinfo{title}{Elements of Applied Bifurcation Theory}}
  (\bibinfo{publisher}{Springer}, \bibinfo{address}{New York, NY},
  \bibinfo{year}{2004}), \bibinfo{edition}{3rd} edn.

\bibitem{AshwinWieczorekVitoloCox}
\bibinfo{author}{Ashwin, P.}, \bibinfo{author}{Wieczorek, S.},
  \bibinfo{author}{Vitolo, R.} \& \bibinfo{author}{Cox, P.}
\newblock \bibinfo{title}{Tipping points in open systems: bifurcation,
  noise-induced and rate-dependent examples in the climate system}.
\newblock \emph{\bibinfo{journal}{Phil. Trans. R. Soc. A}}
  \textbf{\bibinfo{volume}{370}}, \bibinfo{pages}{1166--1184}
  (\bibinfo{year}{2012}).

\bibitem{CarpenterBrock}
\bibinfo{author}{Carpenter, S.} \& \bibinfo{author}{Brock, W.}
\newblock \bibinfo{title}{Rising variance: a leading indicator of ecological
  transition}.
\newblock \emph{\bibinfo{journal}{Ecol. Lett.}}
  \textbf{\bibinfo{volume}{9}}, \bibinfo{pages}{311--318}
  (\bibinfo{year}{2006}).

\bibitem{BoettingerRossHastings}
\bibinfo{author}{Boettinger, C.}, \bibinfo{author}{Ross, N.} \&
  \bibinfo{author}{Hastings, A.}
\newblock \bibinfo{title}{Early warning signals: the charted and uncharted
  territories}.
\newblock \emph{\bibinfo{journal}{Theor. Ecol.}} \textbf{\bibinfo{volume}{6}},
  \bibinfo{pages}{255--264} (\bibinfo{year}{2013}).

\bibitem{LadeGross}
\bibinfo{author}{Lade, S.} \& \bibinfo{author}{Gross, T.}
\newblock \bibinfo{title}{Early warning signals for critical transitions: a
  generalized modeling approach}.
\newblock \emph{\bibinfo{journal}{PLoS Comp. Biol.}}
  \textbf{\bibinfo{volume}{8}}, \bibinfo{pages}{e1002360--6}
  (\bibinfo{year}{2012}).

\bibitem{Veraartetal}
\bibinfo{author}{Veraart, A.} \emph{et~al.}
\newblock \bibinfo{title}{Recovery rates reflect distance to a tipping point in
  a living system}.
\newblock \emph{\bibinfo{journal}{Nature}} \textbf{\bibinfo{volume}{481}},
  \bibinfo{pages}{357--359} (\bibinfo{year}{2012}).

\bibitem{KuehnCT2}
\bibinfo{author}{Kuehn, C.}
\newblock \bibinfo{title}{{A mathematical framework for critical transitions:
  normal forms, variance and applications}}.
\newblock \emph{\bibinfo{journal}{J. Nonlinear Sci.}}
  \textbf{\bibinfo{volume}{23}}, \bibinfo{pages}{457--510}
  (\bibinfo{year}{2013}).

\bibitem{DrakeGriffen}
\bibinfo{author}{Drake, J.} \& \bibinfo{author}{Griffen, B.}
\newblock \bibinfo{title}{Early warning signals of extinction in deteriorating
  environments}.
\newblock \emph{\bibinfo{journal}{Nature}} \textbf{\bibinfo{volume}{467}},
  \bibinfo{pages}{456--459} (\bibinfo{year}{2010}).

\bibitem{DaiVorselenKorolevGore}
\bibinfo{author}{Dai, L.}, \bibinfo{author}{Vorselen, D.},
  \bibinfo{author}{Korolev, K.} \& \bibinfo{author}{Gore, J.}
\newblock \bibinfo{title}{Generic indicators for loss of resilience before a
  tipping point leading to population collapse}.
\newblock \emph{\bibinfo{journal}{Science}} \textbf{\bibinfo{volume}{336}},
  \bibinfo{pages}{1175--1177} (\bibinfo{year}{2012}).

\bibitem{DushoffLevin}
\bibinfo{author}{Dushoff, J.} \& \bibinfo{author}{Levin, S.}
\newblock \bibinfo{title}{The effects of population heterogeneity on disease
  invasion}.
\newblock \emph{\bibinfo{journal}{Math. Biosci.}}
  \textbf{\bibinfo{volume}{128}}, \bibinfo{pages}{25--40}
  (\bibinfo{year}{1995}).

\bibitem{CushingDennisDesharnaisCostantino}
\bibinfo{author}{Cushing, J.}, \bibinfo{author}{Dennis, B.},
  \bibinfo{author}{Desharnais, R.} \& \bibinfo{author}{Costantino, R.}
\newblock \bibinfo{title}{Moving toward an unstable equilibrium: saddle nodes
  in population systems}.
\newblock \emph{\bibinfo{journal}{J. Anim. Ecol.}}
  \textbf{\bibinfo{volume}{67}}, \bibinfo{pages}{298--306}
  (\bibinfo{year}{1998}).

\bibitem{Hastings7}
\bibinfo{author}{Hastings, A.}
\newblock \bibinfo{title}{Transients: the key to long-term ecological
  understanding?}
\newblock \emph{\bibinfo{journal}{Trends Ecol. Evol.}}
  \textbf{\bibinfo{volume}{19}}, \bibinfo{pages}{39--45}
  (\bibinfo{year}{2004}).

\bibitem{May1}
\bibinfo{author}{May, R.}
\newblock \bibinfo{title}{Will a large complex system be stable?}
\newblock \emph{\bibinfo{journal}{Nature}} \textbf{\bibinfo{volume}{238}},
  \bibinfo{pages}{413--414} (\bibinfo{year}{1972}).

\bibitem{Izhikevich}
\bibinfo{author}{Izhikevich, E.}
\newblock \bibinfo{title}{Neural excitability, spiking, and bursting}.
\newblock \emph{\bibinfo{journal}{Int. J. Bif. Chaos}}
  \textbf{\bibinfo{volume}{10}}, \bibinfo{pages}{1171--1266}
  (\bibinfo{year}{2000}).\quad

\bibitem{AshwinField}
\bibinfo{author}{Ashwin, P.} \& \bibinfo{author}{Field, M.}
\newblock \bibinfo{title}{Heteroclinic networks in coupled cell systems}.
\newblock \emph{\bibinfo{journal}{Arch. Rat. Mech. Anal.}}
  \textbf{\bibinfo{volume}{148}}, \bibinfo{pages}{107--143}
  (\bibinfo{year}{1999}).

\bibitem{HuismanWeissing}
\bibinfo{author}{Huisman, J.} \& \bibinfo{author}{Weissing, F.}
\newblock \bibinfo{title}{Biological conditions for oscillations and chaos
  generated by multispecies competition}.
\newblock \emph{\bibinfo{journal}{Ecology}} \textbf{\bibinfo{volume}{82}},
  \bibinfo{pages}{2682--2695} (\bibinfo{year}{2001}).

\bibitem{LiebholdBascompte}
\bibinfo{author}{Liebhold, A.} \& \bibinfo{author}{Bascompte, J.}
\newblock \bibinfo{title}{The Allee effect, stochastic dynamics and the eradication of alien species}.
\newblock \emph{\bibinfo{journal}{Ecol. Lett.}} \textbf{\bibinfo{volume}{6}},
  \bibinfo{pages}{133--140} (\bibinfo{year}{2003}).

\bibitem{ZschalerTraulsenGross}
\bibinfo{author}{Zschaler, G.}, \bibinfo{author}{Traulsen, A.} \&
  \bibinfo{author}{Gross, T.}
\newblock \bibinfo{title}{A homoclinic route to asymptotic full cooperation in
  adaptive networks and its failure}.
\newblock \emph{\bibinfo{journal}{New J. Phys.}}
  \textbf{\bibinfo{volume}{12}}, \bibinfo{pages}{(093015)}; 
	DOI:10.1088/1367-2630/12/9/093015 (\bibinfo{year}{2010}).

\bibitem{DoebliHauert}
\bibinfo{author}{Doebeli, M.} \& \bibinfo{author}{Hauert, C.}
\newblock \bibinfo{title}{{Models of cooperation based on the Prisoner's
  Dilemma and the Snowdrift game}}.
\newblock \emph{\bibinfo{journal}{Ecol. Lett.}} \textbf{\bibinfo{volume}{8}},
  \bibinfo{pages}{748--766} (\bibinfo{year}{2005}).

\bibitem{GrossDLimaBlasius}
\bibinfo{author}{Gross, T.}, \bibinfo{author}{D'Lima, C.~D.} \&
  \bibinfo{author}{Blasius, B.}
\newblock \bibinfo{title}{Epidemic dynamics on an adaptive network}.
\newblock \emph{\bibinfo{journal}{Phys. Rev. Lett.}}
  \textbf{\bibinfo{volume}{96}}, \bibinfo{pages}{(208701)}
  (\bibinfo{year}{2006}).

\bibitem{GrossKevrekidis}
\bibinfo{author}{Gross, T.} \& \bibinfo{author}{Kevrekidis, I.}
\newblock \bibinfo{title}{Robust oscillations in {SIS} epidemics on adpative
  networks: coarse-graining by automated moment closure}.
\newblock \emph{\bibinfo{journal}{Europhys. Lett.}}
  \textbf{\bibinfo{volume}{82}}, \bibinfo{pages}{(38004)}
  (\bibinfo{year}{2008}).

\bibitem{OReganDrake}
\bibinfo{author}{O'Regan, S.} \& \bibinfo{author}{Drake, J.}
\newblock \bibinfo{title}{Theory of early warning signals of disease emergence
  and leading indicators of elimination}.
\newblock \emph{\bibinfo{journal}{Theor. Ecol.}} \textbf{\bibinfo{volume}{6}},
  \bibinfo{pages}{333--357} (\bibinfo{year}{2013}).

\bibitem{BjornstadFinkenstaedtGrenfell}
\bibinfo{author}{Bj{\o}rnstad, O.}, \bibinfo{author}{Finkenst{\"{a}}dt, B.} \&
  \bibinfo{author}{Grenfell, B.}
\newblock \bibinfo{title}{{Endemic and epidemic dynamics of measles. I.
  Estimating epidemiological scaling with a time series SIR model}}.
\newblock \emph{\bibinfo{journal}{Ecol. Monogr.}}
  \textbf{\bibinfo{volume}{72}}, \bibinfo{pages}{169--184}
  (\bibinfo{year}{2002}).

\bibitem{GrenfellBjornstadFinkenstaedt}
\bibinfo{author}{Grenfell, B.}, \bibinfo{author}{Bj{\o}rnstad, O.} \&
  \bibinfo{author}{Finkenst{\"{a}}dt, B.}
\newblock \bibinfo{title}{{Endemic and epidemic dynamics of measles. II.
  Scaling predictability, noise and determinism with the time-series SIR
  model}}.
\newblock \emph{\bibinfo{journal}{Ecol. Monogr.}}
  \textbf{\bibinfo{volume}{72}}, \bibinfo{pages}{185--202}
  (\bibinfo{year}{2002}).

\bibitem{HallerbergKantz}
\bibinfo{author}{Hallerberg, S.} \& \bibinfo{author}{Kantz, H.}
\newblock \bibinfo{title}{Influence of the event magnitude on the
  predictability of extreme events}.
\newblock \emph{\bibinfo{journal}{Phys. Rev. E}} \textbf{\bibinfo{volume}{77}},
  \bibinfo{pages}{011108} (\bibinfo{year}{2008}).

\bibitem{BoettingerHastings}
\bibinfo{author}{Boettinger, C.} \& \bibinfo{author}{Hastings, A.}
\newblock \bibinfo{title}{Quantifying limits to detection of early warning for
  critical transitions}.
\newblock \emph{\bibinfo{journal}{J. R. Soc. Interface}}
  \textbf{\bibinfo{volume}{9}}, \bibinfo{pages}{2527--2539}
  (\bibinfo{year}{2012}).

\bibitem{DerrickVandenDriessche}
\bibinfo{author}{Derrick, W.} \& \bibinfo{author}{den Driessche, P.~V.}
\newblock \bibinfo{title}{Homoclinic orbits in a disease transmission model
  with nonlinear incidence and nonconstant population}.
\newblock \emph{\bibinfo{journal}{Discr. Cont. Dyn. Syst. B}}
  \textbf{\bibinfo{volume}{3}}, \bibinfo{pages}{299--311}
  (\bibinfo{year}{2003}).

\bibitem{VandenDriesscheWatmough}
\bibinfo{author}{den Driessche, P.~V.} \& \bibinfo{author}{Watmough, J.}
\newblock \bibinfo{title}{Epidemic solutions and endemic catastrophes}.
\newblock In \bibinfo{editor}{Ruan, S.}, \bibinfo{editor}{Wolkowicz, G.} \&
  \bibinfo{editor}{Wu, J.} (eds.) \emph{\bibinfo{booktitle}{Dynamical Systems
  and Their Applications in Biology}}, \bibinfo{pages}{247--257}
  (\bibinfo{publisher}{AMS}, \bibinfo{address}{Providence,
  USA}, \bibinfo{year}{2003}).

\bibitem{RohaniKeelingGrenfell}
\bibinfo{author}{Rohani, P.}, \bibinfo{author}{Keeling, M.} \&
  \bibinfo{author}{Grenfell, B.}
\newblock \bibinfo{title}{The interplay between determinism and stochasticity
  in childhood diseases}.
\newblock \emph{\bibinfo{journal}{Amer. Nat.}} \textbf{\bibinfo{volume}{159}},
  \bibinfo{pages}{469--481} (\bibinfo{year}{2002}).

\bibitem{FinkenstaedtGrenfell}
\bibinfo{author}{Finkenst{\"a}dt, B.} \& \bibinfo{author}{Grenfell, B.}
\newblock \bibinfo{title}{Time series modelling of childhood diseases: a
  dynamical systems approach}.
\newblock \emph{\bibinfo{journal}{J. R. Stat. Soc. C}}
  \textbf{\bibinfo{volume}{49}}, \bibinfo{pages}{187--205}
  (\bibinfo{year}{2000}).

\bibitem{Bright2}
\bibinfo{author}{Bright, C.}
\newblock \emph{\bibinfo{title}{Life out of Bounds: Bioinvasion in a Borderless
  World}} (\bibinfo{publisher}{W. W. Norton}, \bibinfo{address}{New York City,
  USA}, \bibinfo{year}{1998}).

\bibitem{PhilippsBrownWebbShine}
\bibinfo{author}{Phillips, B.}, \bibinfo{author}{Brown, G.},
  \bibinfo{author}{Webb, J.} \& \bibinfo{author}{Shine, R.}
\newblock \bibinfo{title}{Invasion and the evolution of speed in toads}.
\newblock \emph{\bibinfo{journal}{Nature}} \textbf{\bibinfo{volume}{439}},
  \bibinfo{pages}{803} (\bibinfo{year}{2006}).

\bibitem{TaoVu}
\bibinfo{author}{Tao, T.} \& \bibinfo{author}{Vu, V.}
\newblock \bibinfo{title}{Random matrices: the circular law}.
\newblock \emph{\bibinfo{journal}{Commun. Contemp. Math.}}
  \textbf{\bibinfo{volume}{10}}, \bibinfo{pages}{261--307}
  (\bibinfo{year}{2008}).

\bibitem{GardnerAshby}
\bibinfo{author}{Gardner, M.} \& \bibinfo{author}{Ashby, W.}
\newblock \bibinfo{title}{Connectance of large dynamic (cybernetic) systems:
  critical values for stability}.
\newblock \emph{\bibinfo{journal}{Nature}} \textbf{\bibinfo{volume}{228}},
  \bibinfo{pages}{784} (\bibinfo{year}{1970}).

\bibitem{WainribTouboul}
\bibinfo{author}{Wainrib, G.} \& \bibinfo{author}{Touboul, J.}
\newblock \bibinfo{title}{Topological and dynamical complexity of random neural
  networks}.
\newblock \emph{\bibinfo{journal}{Phys. Rev. Lett.}}
  \textbf{\bibinfo{volume}{110}}, \bibinfo{pages}{118101}
  (\bibinfo{year}{2013}).

\bibitem{Gardiner}
\bibinfo{author}{Gardiner, C.}
\newblock \emph{\bibinfo{title}{Stochastic Methods}}
  (\bibinfo{publisher}{Springer}, \bibinfo{address}{Berlin Heidelberg,
  Germany}, \bibinfo{year}{2009}), \bibinfo{edition}{4th} edn.

\bibitem{LiuLevinIwasa}
\bibinfo{author}{Liu, W.-M.} \& \bibinfo{author}{Levin, S.A.} \& \bibinfo{author}{Iwasa, Y.}
\newblock \bibinfo{title}{Influence of nonlinear incidence rates upon the be-
haviour of {SIRS} epidemiological models}.
\newblock \emph{\bibinfo{journal}{J. Math. Biol.}}
  \textbf{\bibinfo{volume}{23}}, \bibinfo{pages}{187--204}
  (\bibinfo{year}{1986}).

\bibitem{BrauervandenDriesscheWu}
\bibinfo{author}{Brauer, F.}, \bibinfo{author}{van~den Driessche, P.} \&
  \bibinfo{author}{Wu, J.}
\newblock \emph{\bibinfo{title}{Mathematical Epidemiology}}
  (\bibinfo{publisher}{Springer}, \bibinfo{address}{Berlin Heidelberg,
  Germany}, \bibinfo{year}{2008}).

\bibitem{GrenfellWeb}
\bibinfo{author}{Grenfell, B.}
\newblock TSIR analysis of measles in England and Wales
  \bibinfo{note}{{http://www.zoo.cam.ac.uk/zoostaff/grenfell/measles.htm}; Date of access:24/11/2011}.

\bibitem{GrenfellWeb1}
\bibinfo{author}{Grenfell, B.}
\newblock TSIR analysis of measles in England and Wales (archived version of webpage)
  \bibinfo{note}{{http://www.asc.tuwien.ac.at/$\sim$ckuehn/Data/DATA\_measles.zip}; 
  Date of access:27/10/2014}.  

\end{thebibliography}
\end{document}